\documentclass[notoc,12pt]{JHEP3}

\input{epsf}

\usepackage{epsfig}

\title{The scaling region of the lattice $O(N)$ sigma model at finite temperature}
\author{Costas G. Strouthos$^a$ and Ioannis N. Tziligakis$^b$ \\
$^a\,${\it Department of Physics, University of Wales Swansea,\\
Singleton Park, Swansea SA2 8PP, U.K.}\\
$^b\,${\it Department of Physics, University of Illinois at Urbana-Champaign,\\
Urbana, Illinois 61801-3080, U.S.A.}\\
}
\abstract{ 
We present results from numerical studies of the
finite temperature phase transition of the $(3+1)d$ 
$O(N)$-symmetric
non-linear sigma model for $N=1,2$ and $3$. 
We study the dependence of the width of the $3d$ critical region on $N$
and we show that the broken phase scaling region is much wider for $N=2$ and $3$ 
than for $N=1$. 
We also compare the widths of the critical region 
in the low $T$ and high $T$ phases of the $O(2)$ model and we show that 
the scaling region in the broken phase is much wider than in the symmetric phase. 
We also report results for the width of the scaling regions in the low $T$ phase
$(2+1)d$ Ising model and we show that the spatial correlation length has to be
approximately twice the lattice temporal extent before the $2d$ scaling region is reached.
}

\preprint{SWAT/02/364}

\begin{document}

\section{Introduction}
\label{intro}

The behavior of symmetries at finite temperature is one of the most outstanding
and relevant problems in current areas of particle physics such as cosmology,
relativistic heavy-ion collisions and the quark-gluon plasma. 
It was shown some time ago that spontaneously broken symmetries in quantum 
field theories get restored at sufficiently high temperature \cite{linde}. 
In the high temperature limit, however, perturbation theory becomes unreliable 
when infrared divergences are not properly summed. Various groups
developed temperature dependent renormalization group techniques to study 
the critical properties of the $O(N)$ scalar field theories (for a few examples see \cite{analytical}
and references therein).

In recent years considerable work has been done on the  physics of the finite temperature
chiral phase transition in QCD and related models. A compelling idea 
based on universality and dimensional reduction first put forward in 
\cite{wilczek}, is that in QCD with two massless quarks the physics near the transition 
can be described by the three-dimensional $O(4)$-symmetric sigma model.  
Although, there is little disagreement that the chiral phase transition in QCD 
is second order, no quantitative work or simulations have been done that decisively 
determine its universality class, because simulations are performed away from the 
continuum limit on small 
volumes and show substantial discretization and finite size effects. 
The knowledge of the scaling relations are crucial for
a successful extrapolation to small mass and large volume near $T_c$.

The reasoning behind the dimensional reduction scenario is based on the dominance of the light degrees
of freedom near $T_c$. In the imaginary time formalism of finite temperature field theory
a given system is defined on a manifold $S^1 \times R^d$ with the inverse temperature 
$\beta \equiv T^{-1}$ being the circumference of $S^1$. The boundary condition in the 
time direction is  periodic for bosons and anti-periodic for fermions. At temperatures 
larger than the physical scales of the system, the nonzero Matsubara modes acquire 
a heavy mass and non-static configurations are strongly suppressed in the Boltzmann
sum. If the original four-dimensional theory has fermions, then because of the anti-periodicity 
in the temporal direction, all fermionic modes become massive and can be integrated over. 
This is not to say that non-static modes have no effect. Rather they generate counter-terms
to the three-dimensional theory. 
The physical picture of the dimensional reduction scenario can be described 
nicely by using a slightly
different language presented in \cite{stephanov}.
As one approaches the critical point, the fluctuations in space become correlated over a distance 
$\xi$ which diverges at the critical point. In the critical region, not only do the fluctuations
extend over large spatial scales, they also become very slow and there is a divergent relaxation time
$\tau$; the effect is known as critical slowing down. The appropriate quantum of energy 
is then $\hbar \omega_c$ with $\omega_c \sim 1/\tau$. Hence if the transition is at a finite
temperature $T_c$, $\hbar \omega_c \ll k_BT$ near the critical point, which means that long 
wavelength fluctuations are classical thermal fluctuations and quantum mechanics is irrelevant. 
Equivalently, since the time interval $\beta$ is very short compared to the relaxation time
$\tau$, typical time histories will consist of a single configuration which is the 
same at each time slice. The dynamics therefore drops out of the problem and a Boltzmann
weight $\exp(-\beta H_{\rm classical})$ is recovered from the path integral. 

In this work we study numerically the finite temperature transition both in the $(3+1)d$ 
and $(2+1)d$ $\phi^4$ model. 
This is equivalent to studying the phenomenon of dimensional crossover in a $d$-dimensional
layer $O(N)$ spin model of infinite extent in $d-1$ dimensions and finite 
thickness in the remaining dimension. In other words temperature effects become 
predominant as a consequence of a change in dimension.
The formal derivations of the dimensional reduction scenario require that the scalar mass is 
small compared to the temperature. We check whether this condition is necessary and we study 
the effects of the infrared fluctuations on the width of the critical region under different 
circumstances.
We performed simulations of the model with different $O(N)$ (for $N=1,2$ and $3$) symmetries in $(3+1)d$ 
and study the dependence of the width of the $3d$ scaling region on $N$.
In the case of the $(3+1)d$ $O(2)$ model we compare the widths of the scaling region in the 
low temperature broken phase 
and the high temperature symmetric phase.
We also study the scaling region of the $(2+1)d$ $O(1) \equiv Z_2$ model.
In addition we are interested to see how finite size
effects can distort the results in such a simple model that can be simulated
on lattices with relatively large temporal and spatial sizes. 
We extract critical exponents by measuring   
the order parameter, the susceptibility, and the correlation length.

\section{The model}
\label{Formulation}
In the continuum limit the Lagrangian of the model reads:
\begin{equation}
{\cal L} = \frac{1}{2}(\partial \vec{\phi})^2+ \frac{m_0^2}{2}\vec{\phi}^2
+\frac{g}{4!}(\vec{\phi}^2)^2.
\end{equation}
In order to study the problem of symmetry restoration in a non-perturbative
way, we regard this model as an effective low energy theory by introducing a lattice
regularization with a lattice constant $a = 1/\Lambda$, where $\Lambda$ is the ultraviolet
cut-off.
Summing the bare Lagrangian 
over all lattice sites and transforming the continuum parameters into 
suitably defined lattice parameters according to a standard convention
\begin{equation}
a\vec{\phi}(x)=\kappa^{1/2}\vec{\Phi}_x, \;\;\;\;\; a^2m_0^2 = \frac{2-4\lambda}{\kappa}-8, 
\;\;\;\; g=\frac{24\lambda}{\kappa^2},
\end{equation}
we are led to the following action:
\begin{equation}
S[\vec{\Phi}]=\sum_x\left[-\kappa \sum_{\mu}\Phi_x^{\alpha}\Phi_{x+\mu}^{\alpha}
+\Phi_x^{\alpha}\Phi_x^{\alpha}+\lambda(\Phi_x^{\alpha}\Phi_x^{\alpha}-1)^2 \right],
\end{equation}
where the index $\mu$ runs over the nearest neighbors, ($\mu=1,...,d$), 
and the index $\alpha$ runs over the $N$ components of the field $\vec{\Phi}$. 
For convenience, in our numerical work we concentrated on the limit of an infinitely 
strong bare coupling where the field $\vec{\Phi}$ is constrained to unit length:
$\vec{\Phi}\cdot\vec{\Phi}=1$. In this case the action $S$ reduces to the form:
\begin{equation}
S[\Phi]=-\kappa \sum_{x,\mu}\Phi_x^{\alpha}\Phi_{x+\mu}^{\alpha},
\end{equation} 
which is equivalent to the energy of a configuration of the classical ferromagnetic 
$O(N)$ spin model.

Finite physical temperature on the lattice means a finite size of the temporal extent
$L_t$. The physical temperature is given by $T=1/(L_ta)$. 
The temperature of the system may be altered, by varying either the coupling 
$\kappa$, which amounts to varying the lattice spacing or by varying $L_t$.
To minimize the effects of lattice artifacts the condition $\Lambda \gg T$ must be satisfied
which requires a lattice temporal extent $L_t \gg 1$. This condition not being met 
implies that the sum over the Matsubara modes is truncated at a small value.
It is well-known that the $(3+1)d$ renormalized $\phi^4$ theory is a free theory.
However, the regularized theory may well describe the correct physics of a broad
spectrum of processes below the cut-off scale. 

As discussed in \cite{jansen} the regularized model can be interpreted in terms of 
physical terms by moving along the curves of constant physics (CCPs) which are 
characterized by $g_R = {\rm const.}$ Therefore, when approaching the $T=0$
second order critical line the mass of the sigma meson $M_{\sigma}$ decreases 
and in order to keep $T/M_{\sigma}$ 
constant one has to increase $L_t$. Near the $T=0$ critical line the ratio $T/M_{\sigma}$
should not change any more along a CCP. It was shown within a specific regularization 
scheme to leading order in $1/N$ \cite{bardeen}
that $T_c/M_{\sigma} = \sqrt{36/(Ng_R)}$ and by solving the gap equation on the lattice it was
shown \cite{jansen} that the corrections due to a finite $L_t$ are negligible for $L_t \geq 4$.
In order to suppress possible discretization effects arising from $1/N$ corrections we performed 
our simulations with $L_t=12$. The $(2+1)d$ $\phi^4$ theory has an interacting continuum limit and
it has been shown that the non-linear sigma model is renormalizable in the $1/N$
expansion \cite{arefeva}. The phenomenon of symmetry restoration at finite temperature 
in the three-dimensional theory was investigated by means of lattice simulations
and finite size scaling techniques by Bimonte et al. \cite{bimonte}. The authors of \cite{bimonte} located 
the intersections of critical lines corresponding to different values of $L_t$ with various CCPs  
lying in the ordered phase of the $T=0$ theory and showed
that for $L_t \geq 3$ the critical temperatures approach the finite physical temperature of the 
continuum theory. 

Furthermore, the condition $L_s \gg \xi$ must be satisfied in order to obtain results
that approximate the thermodynamic limit. For this reason at certain values of coupling $\kappa$
we performed simulations on lattices with different spatial extent in order to 
detect and control finite size effects.

In our simulations we used the single-cluster algorithm \cite{wolff},
because it reduces dramatically the correlations
between successive configurations near a critical point. The cluster update is as follows: 
First choose randomly a reflection axis in the plane, denote the component of the spin 
$\vec{\Phi}_i$ that is parallel to this reflection axis by $\Phi^{||}_i$ and that which 
is orthogonal by $\Phi^{\perp}_i$. Then choose randomly a site $i$ of the lattice as a starting 
point for the cluster ${\cal C}$. Visit all neighbor sites $j$ of $i$ and allow them to 
join the cluster with probability
\begin{equation}
p(i,j)=1-\exp[-\kappa(\Phi^{\perp}_i \Phi^{\perp}_j + |\Phi^{\perp}_i\Phi^{\perp}_j|)].
\end{equation}
After this is done, visit all neighbors of the new sites in the cluster and add them 
to the cluster with probability $p(i,j)$. Iterate this step until no new sites enter the
cluster. Finally flip the signs of all $\Phi^{\perp}_i$ contained in the cluster.

\section{The observables}
\label{observables}
In this section we present the various observables measured in the broken and symmetric  
phases of the model in our lattice simulations.
The magnetization $M^{\alpha}$ of a given configuration is defined as:
\begin{equation}
M^{\alpha} \equiv \frac{1}{V}\sum_x \Phi^{\alpha}_x,
\end{equation}
where $V \equiv L_s^3 L_t$ is the space-time volume. However, in a finite volume and without the 
benefit of an explicit symmetry breaking interaction with an external magnetic field, the 
direction of symmetry 
breaking changes over the course of the run so that $M^{\alpha}$ averages to zero over 
the ensemble. It is in this way that the absence of spontaneous symmetry breaking on a finite
lattice is enforced. Another option is to introduce a field operator 
\begin{equation}
\Phi_{\sigma,x} \equiv \frac{\Phi_x^{\alpha}M^{\alpha}}{|M|},
\end{equation}
which is a projection of $\Phi_x^{\alpha}$ to the direction of $M^{\alpha}$ 
separately for each configuration \cite{hasenfratz89}. The effective order parameter 
\begin{equation}
\Sigma \equiv \frac{1}{V}\langle\sum_{x}\Phi_{\sigma,x}\rangle=\langle|M|\rangle,
\end{equation}
differs
from the true order parameter extrapolated to the zero magnetic field limit by a factor $V^{-\frac{1}{2}}$
\cite{hasenfratz90}.
One can easily show that the susceptibility of the order parameter can be expressed in terms of moments
of the magnetization as follows:
\begin{equation}
\chi=V(\langle|M|^2\rangle - \langle |M| \rangle^2).
\end{equation}
In the symmetric phase one gets $\chi=V\langle|M|^2\rangle$.
Cluster algorithms enable us to reduce the variance of the expectation values
in the symmetric phase by using improved estimators \cite{wolff,hasenbusch90}.
The improved estimator for the susceptibility is given by:
\begin{equation}
\label{suscimp}
\chi_{\rm imp}=\left\langle \frac{N}{|C|}\left( \sum_{i \in {\cal C}} 
\Phi_i^{\perp} \right)^2 \right \rangle, 
\end{equation}
where $|C|$ denotes the number of spins in the cluster ${\cal C}$.

We determined the temporal exponential correlation length $\xi_{\rm exp}^t$ of the sigma meson from the 
asymptotic exponential decay of the zero momentum connected two-point correlation function constructed 
from the field $\Phi_{\sigma,x}$, i.e.,
\begin{eqnarray}
\label{correl}
O(t)& = & \sum_{\vec{x}} \Phi_{\sigma,x} \\
G(t)& \equiv & \langle O_0 O_t \rangle - \Sigma^2 \stackrel{t \gg1}{=} A 
\left[\exp\left(\frac{-t}{\xi_{\rm exp}}\right)
+\exp\left(\frac{-(L_t-1)}{\xi_{\rm exp}}\right) \right]. 
\end{eqnarray}
Similarly, we measured the spatial correlation length $\xi_{\rm exp}^s$ from the exponential 
decay of $G(x)$ in the spatial direction. 
The asymmetry of the lattice at nonzero temperature implies that 
$\xi_{\rm exp}^s \neq \xi_{\rm exp}^t$, 
because temperature effects on the mesonic self-energy become
direction dependent. The breaking of Lorentz invariance implies that the spectral
functions and the dispersion relations are modified in a non-trivial way 
by the thermal statistical ensemble.
However, both $\xi_{\rm exp}^s$ and $\xi_{\rm exp}^t$ are expected to 
diverge as $T \to T_c$ according to the scaling law $\sim (T_c-T)^{-\nu}$, where $\nu$ 
is a critical exponent of the corresponding $O(N)$ $3d$ spin model.
One can construct correlation functions from
various other operators $O(t)$ that couple to the scalar particle $\sigma$, but 
according to \cite{hasenfratz89} the operators we defined in eq.(\ref{correl})
 give a better signal than other options. 
In the symmetric phase the two-point correlation function is 
$G(t)=\left\langle(\sum_{i}\vec{\Phi}(\vec{x}_i,t)) 
\cdot \sum_{j}(\vec{\Phi}(\vec{x}_j,0))\right\rangle$ and its improved estimator
is given by 
\begin{equation}
\label{xiexpimp}
G(t)_{\rm imp}= \left \langle \frac{N}{|{\cal C}|}\delta_{ij}({\cal C})
\Phi^{\perp}_i \Phi^{\perp}_j \right \rangle, 
\end{equation}
where $\delta_{ij}({\cal C})=1$ if $i$ and $j$ belong to the same cluster otherwise 
$\delta_{ij}({\cal C})=0$ \cite{wolff,hasenbusch90}.

We also measured the so-called second moment
correlation length $\xi_{\rm 2nd}$, defined by 
\begin{equation}
\label{xi2}
\xi_{\rm 2nd}^2 \equiv \frac{1}{2d} \frac{\sum_{x} |x|^2 G(x)}{\sum_{x}G(x)} 
=-\left( \bar{G}^{-1}(k^2) \frac{d}{dk^2}\bar{G}(k^2) \right)_{k^2=0},
\end{equation}
where $\bar{G}(k) \equiv \sum_{j}\langle \exp(ikx_j)\vec{\Phi}_0\vec{\Phi}_j \rangle$
is the Fourier transform of the two-point correlation function.
On the lattice $\xi_{\rm 2nd}$ is expressed as
\begin{equation}
\label{xi2imp}
\xi_{\rm 2nd} = \left(\frac{(\chi/F)-1}{4\sin^2(\pi/L)}\right)^{1/2},
\end{equation}
where $L=L_t$ or $L_s$, $F \equiv \left. \bar{G}(k) \right|_{|k|=2\pi/L}$.
The improved $\xi_{\rm 2nd}$ requires $\bar{G}(k)_{\rm imp}$ given by
\begin{equation}
\bar{G}(k)_{\rm imp} = \left \langle \frac{N}{|{\cal C}|} \left[ \left( \sum_{i \in {\cal C}}
\Phi_i^{\perp} \cos(kx_i) \right)^2 
+ \left(\sum_{i \in {\cal C}} \Phi_i^{\perp} \sin(kx_i) \right)^2 \right] \right \rangle. 
\end{equation}
This estimator of the correlation length is very popular, because its numerical evaluation 
is simpler than that of the exponential correlation length which requires an identification 
of an asymptotic exponential decay. 
However, the two definitions of the correlation length do not coincide, since in $\xi_{\rm exp}$ 
only the ground state enters, while in the case of $\xi_{\rm 2nd}$ according to eq.(\ref{xi2})
a mixture of the full spectrum is taken into account. As mentioned in \cite{hasenbusch97}
the difference from one of the ratio $\xi_{\rm exp}/\xi_{\rm 2nd}$ gives an idea of the 
density of the lowest states of the spectrum. It is well-known however, that in the high temperature
phase the asymptotic exponential decay behavior of $G(t)$ is reached at much smaller values of 
$t$ than in the broken phase (see e.g. \cite{hasenbusch97}). 
This behavior of $G(t)$ implies that the difference
between $\xi_{\rm exp}$ and $\xi_{\rm 2nd}$ in the symmetric phase is smaller than in the broken 
phase. In our simulations we measured $\xi_{\rm 2nd}$ only in the symmetric phase and we observed that 
the difference  between $\xi_{\rm exp}$ and $\xi_{\rm 2nd}$ vanishes within error.

\section{Results}
\label{result}
In this section we present results from numerical simulations 
of the four-dimensional $O(N)$ ($N=1,2$ and $3$) non-linear sigma model at 
$T \neq 0$. We also present results from simulations preformed in the low temperature phase 
of the three-dimensional Ising model. 
By measuring the correlation length $\xi$ at the crossover into the $3d$
scaling region 
we study the dependence of the width of the critical region 
on the symmetry group $O(N)$ 
and on whether the symmetry 
is manifest ($T > T_c$) or broken ($T < T_c$). 
We define the crossover coupling $\kappa_{\rm cross}$ as the value of $\kappa$ 
that if it is included in the fitting window, the extracted exponent $\beta_{\rm mag}$
does not agree within statistical error with the exponent of the $3d$ model $\beta_{\rm mag}^{\rm 3d}$. 
As we show in the next paragraphs, the change in the value of the exponent 
is accompanied by a substantial increase in the $\chi^2/{\rm d.o.f.}$ of the fit, implying that  
the quality of the fit drops rapidly once we enter the crossover region.
We managed to identify the value of $\kappa_{\rm cross}$ with relatively good accuracy,
because as we will see in the following paragraphs the value of $\kappa$ was varied
by small increments at the crossover into the $3d$ scaling region. 
Throughout this work we used a large lattice temporal extent $L_t=12$ in order to minimize 
discretization effects. 

First, we present the results for the $(3+1)d$ Ising model.
For certain values
of $\kappa$ we monitored finite size effects on the order parameter 
by simulating the model on 
lattices with two different values of  spatial extent $L_s$.
For the final analysis we chose
the data sets for which the systematic errors on $\Sigma$
are less than $0.5\%$. More precisely 
for the range of couplings $ 0.15040\leq \kappa \leq 0.15100$ we chose $L_s=48$, for 
$ 0.15010 \leq \kappa \leq0.15030 $ $L_s=72$ and for $0.14993 \leq \kappa \leq 0.150075$ 
$L_s=96$. For each value of $\kappa$ we typically generated $10^6$ 
sweeps of the algorithm.  
The magnetization data was fitted to the form
\begin{equation} 
\Sigma = A(\kappa-\kappa_c)^{\beta_{\rm mag}}.
\end{equation}
The results for  different ranges of $\kappa$ are 
presented in Table \ref{t1}. 
The simple power law fit is very good for data close to the transition, i.e.
for $\kappa \leq 0.15005$. For these fits the exponent $\beta_{\rm mag}$ is
consistent within error with a recent estimate of the critical index of the $3d$ Ising model
$\beta_{\rm mag}^{\rm 3d}=0.3269(6)$ \cite{talapov}.
As shown in Table \ref{t1},
the value of $\beta_{\rm mag}$  deviates from the $3d$ values
and the quality of the fit gets worse
as we include more data points at larger values of $\kappa$. 
Therefore, we conclude that the crossover to the $3d$ universality 
class occurs at $\kappa_{\rm cross} \approx 0.150075$. 
The data and the fitting curve to the data with $\kappa \leq 0.150025$ are shown in Fig.~\ref{4d_ising}.
The statistical errors are smaller than the size
of the symbols and for this reason we don't plot them in the figure.
The data  for the susceptibility and the correlation length for these 
values of the coupling were very 
noisy and did not allow us to measure the exponents $\gamma$ and $\nu$.

\TABLE
{
\centering
\caption{Values of $\beta_{\rm mag}$ for the $(3+1)d$ Ising model extracted from different ranges of $\kappa$.}
\label{t1}       
\setlength{\tabcolsep}{2.0pc}
\begin{tabular}{|llll|}
\hline
range of $\kappa$ & $\beta_{\rm mag}$ & $\kappa_c$ & $\chi^2/{\rm d.o.f.}$ \\
\hline
0.14993 - 0.150025 & 0.32(3) &  0.149867(9) & 0.3 \\
0.14993 - 0.15005  & 0.34(2) &  0.149863(7) & 0.5  \\
0.14993 - 0.150075 & 0.37(1) &  0.149870(5) &  1.1  \\
0.14993 - 0.150100 &  0.37(1)  & 0.149853(5) &  7.3  \\
\hline
\end{tabular}
}

\FIGURE
{
\centering

\includegraphics[width=11cm]{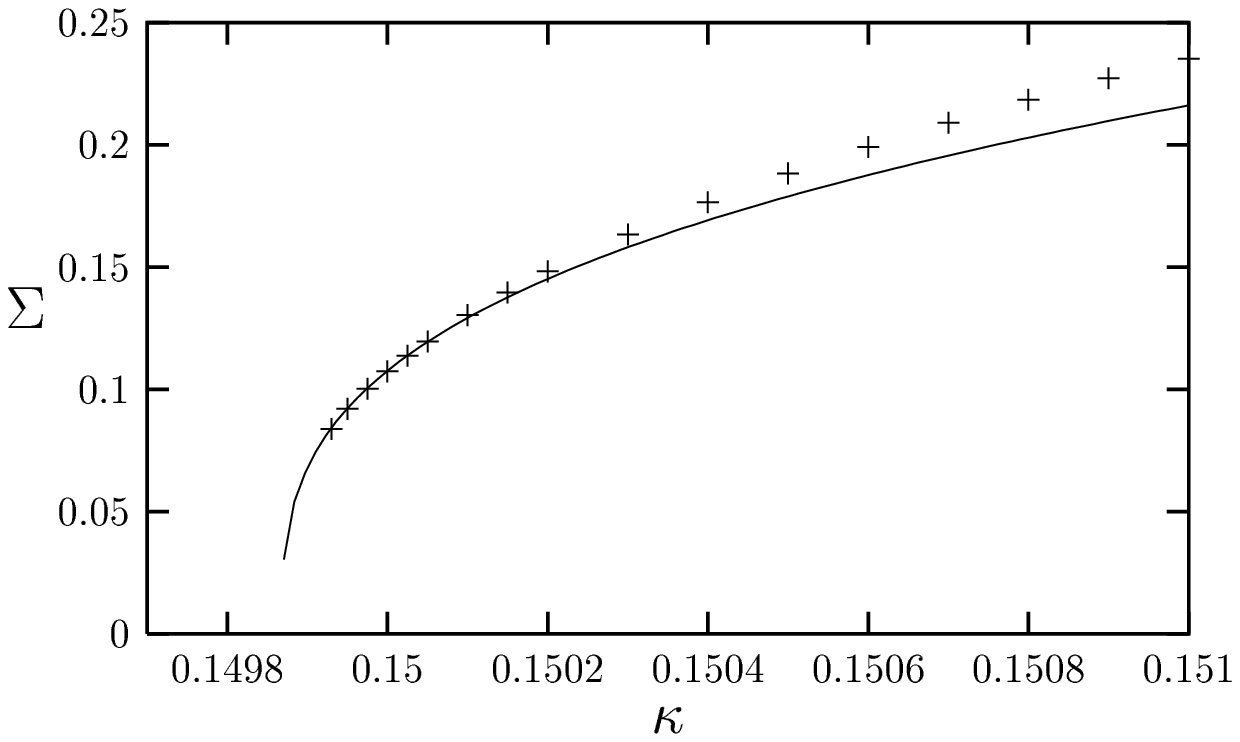}
\caption{Order parameter vs. coupling for the $(3+1)d$ Ising model.
The curve represents the fit to the data for $\kappa \leq 0.15005$.}
\label{4d_ising}
}

We measured the values of the spatial and temporal correlation lengths at the 
crossover into the $3d$ Ising scaling region by performing a simulation 
at  $\kappa=0.150075$
on a $12 \times 48^3$ lattice.  For this measurement we chose a lattice with smaller spatial 
size $L_s$, because the correlation function was very noisy and we had to generate $15$ million 
configurations in order to  measure the correlation lengths $\xi^{t}$ and $\xi^{s}$ with statistical 
errors less than $7\%$.  By fitting the temporal and spatial correlation functions to the single-pole
ansatz we extracted
 $\xi^{t}=6.5(4)$ and $\xi^{s}=8.8(5)$. The overall error was estimated 
from the observation of the fit sensitivity to the variation of the parameters
and to the range of points used in the fit. 
We conclude that in order to enter the $3d$ scaling region
it was not necessary to work so close to the transition itself
that the correlation lengths would be large compared to the temporal extent of the 
lattice. The importance of this comment lies in the fact that the formal derivations
of the dimensional reduction scenario require that the zero mode dominates the Matsubara
sum and this occurs when the scalar mass is small compared to the temperature. Apparently,
our result shows that the higher Matsubara frequencies do not affect the critical singularities
but they do affect non-universal aspects of the transition such as the width of the scaling 
region. 

As the temperature approaches $T_c$ the mass of the sigma meson decreases to zero 
and this results in an increase of the mesonic number density. 
A consequence of this is that the spectral function may become broader due to collisional broadening. 
One can study in detail the properties of the mesons 
at $T \neq 0$ by using the maximum entropy method to extract their spectral functions from high statistics
correlation functions 
measured on anisotropic lattices with temporal sizes larger than $L_t=12$. 
However, for our purposes the $L_t=12$ temporal extent  
is large enough for an estimate of $\xi^{t}$ at $\kappa_{\rm cross}$.
To the extent that we recognized particle-like excitations at $T \neq 0$ 
and $\xi^{t} \neq \xi^{s}$, implies that at $\kappa_{\rm cross}$ the breaking 
of Lorentz invariance caused a significant ``renormalization'' of the speed of light
in the mesonic dispersion relation. 

\TABLE
{
\centering
\caption{Values of $\beta_{\rm mag}$ for the $(3+1)d$ O(2) model extracted from different ranges of $\kappa$.}
\label{t2}       
\setlength{\tabcolsep}{1.7pc}
\begin{tabular}{|lllll|}
\hline
range of $\kappa$ & $L_s$ & $\beta_{\rm mag}$ & $\kappa_c$ &  $\chi^2/{\rm d.o.f}$ \\
\hline
0.3125 - 0.3275 & 24   & 0.340(1) & 0.30358(7)  & 0.5 \\
0.3125 - 0.3275 & 48   & 0.343(2) & 0.30362(1)  &  1.1  \\
0.3125 - 0.3325 & 48   & 0.335(1) &  0.30390(6) &   5.6  \\
0.3125 - 0.3375 & 48  &  0.329(1)   &  0.30417(4) &   7.8  \\
\hline
\end{tabular}
}

\FIGURE
{
\centering
\includegraphics[width=11cm]{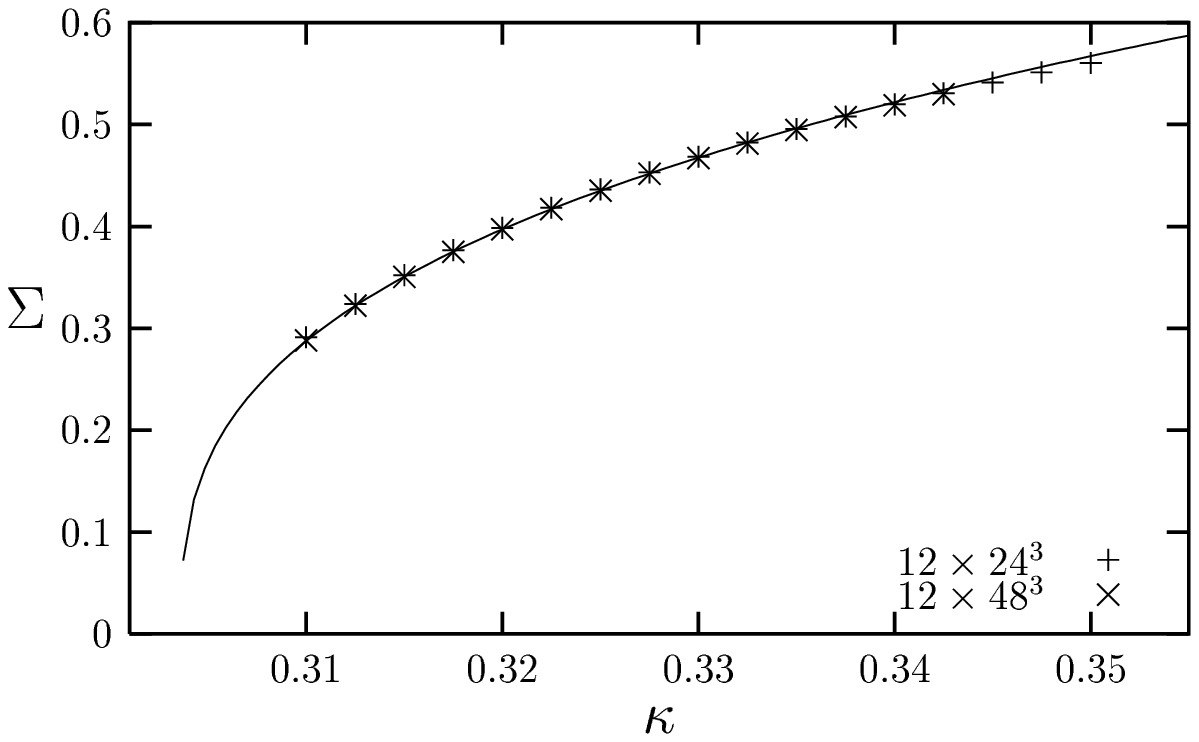}
\caption{Order parameter vs. coupling for the $(3+1)d$ $O(2)$ model measured
on $12 \times 24^3$ and $12 \times 48^3$ lattices.
The curve represents the power law fit to the data for $\kappa \leq 0.3275$.}
\label{order.param_O2}
}

In this paragraph we discuss the numerical results in the low temperature phase of the
O(2) symmetric sigma model.
We generated approximately $80,000$ configurations for each value of the coupling
on the $12 \times 24^3$ lattice and approximately $40,000$ configurations for each value 
of $\kappa$  
on the $12 \times 48^3$ lattice.
The results for the fits of $\Sigma$ to the standard scaling law for 
different values of $\kappa$ are shown in Table \ref{t2}. 
By monitoring the change in the quality of the fits and the deviation in the value 
of $\beta_{\rm mag}$ from recent estimates
of $\beta_{\rm mag}^{\rm 3d}=0.3485(2)$ \cite{hasenbusch01} 
we conclude that $\kappa_{\rm cross} \approx 0.3325$.
An interesting observation is that finite size effects are negligible even 
when the asymmetry ratio is two.   
In Fig.~\ref{order.param_O2} we plot the order parameter versus coupling. The continuous line 
represents the power law fit to the $12 \times 48^3$ data for $\kappa \leq 0.3275$ . 

The values of the spatial and temporal correlation lengths measured at the crossover coupling 
$\kappa_{\rm cross} = 0.3325$ are $\xi^{s}_{\rm cross}=0.94(2)$ and 
$\xi^{t}_{\rm cross}=0.97(3)$. 
We conclude that in the $O(2)$ model the crossover to the $3d$ universality class
occurs at a much smaller value of $\xi/L_t$ than in the 
$O(1)$ model. This explains why the critical region in this case is more easily accessible
on lattices with a smaller asymmetry ratio
than in the case of the Ising model.
The fact that the value of $\xi^{t}_{\rm cross}$ is very close to the value of 
$\xi^{s}_{\rm cross}$ implies that 
the thermal fluctuations are relatively small at the crossover into the $3d$
scaling region.
The diverging correlation length of the Goldstone boson is a source 
of additional finite size effects in a system. 
In this case however, 
the values of the order parameter in the $3d$ scaling region measured 
on small lattices are close to its thermodynamic limit values, 
because the Goldstone boson had the effect of increasing the 
width of the critical region.

\FIGURE
{
\centering
\includegraphics[width=11cm]{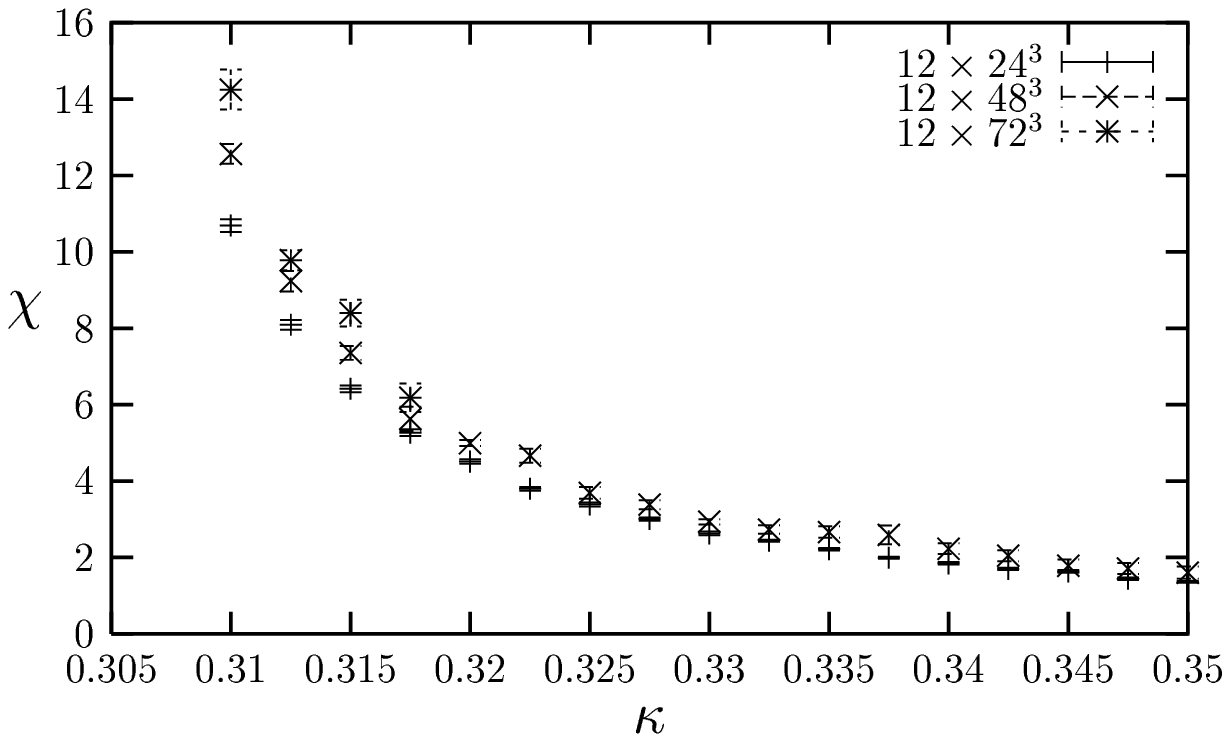}
\caption{Susceptibility vs. coupling for the $(3+1)d$ $O(2)$ model
measured on $12 \times 24^3$, $12 \times 48^3$ and $12 \times 72^3$ lattices.}
\label{susc_O2_broken}
}

In Fig.~\ref{susc_O2_broken} we present the results for the susceptibility $\chi$ of the
order parameter for the $O(2)$ model. As expected finite size effects and statistical
errors are much bigger for $\chi$ than for $\Sigma$ and for this reason we are
unable to extract a meaningful value for the exponent $\gamma$.

\TABLE
{
\centering
\caption{Values of $\beta_{\rm mag}$ for the $(3+1)d$ O(3) model extracted from different ranges of $\kappa$.}
\label{t3}       
\setlength{\tabcolsep}{2.0pc}
\begin{tabular}{|lllll|}
\hline
range of $\kappa$ & $L_s$ & $\beta_{\rm mag}$ & $\kappa_c$ &  $\chi^2/{\rm d.o.f}$ \\
\hline
0.4725 - 0.4875 & 48   & 0.377(4) &  0.4569(2)  &    2.5 \\
0.4725 - 0.4875 & 72   & 0.364(4) &  0.4574(2)  &    0.4  \\
0.4725 - 0.4975 & 72   & 0.356(2) &  0.4579(1)  &    3.5  \\
0.4725 - 0.4500 & 72   & 0.351(1) &  0.4580(1)  &    4.0  \\
\hline
\end{tabular}
}

\FIGURE
{
\centering
\includegraphics[width=11cm]{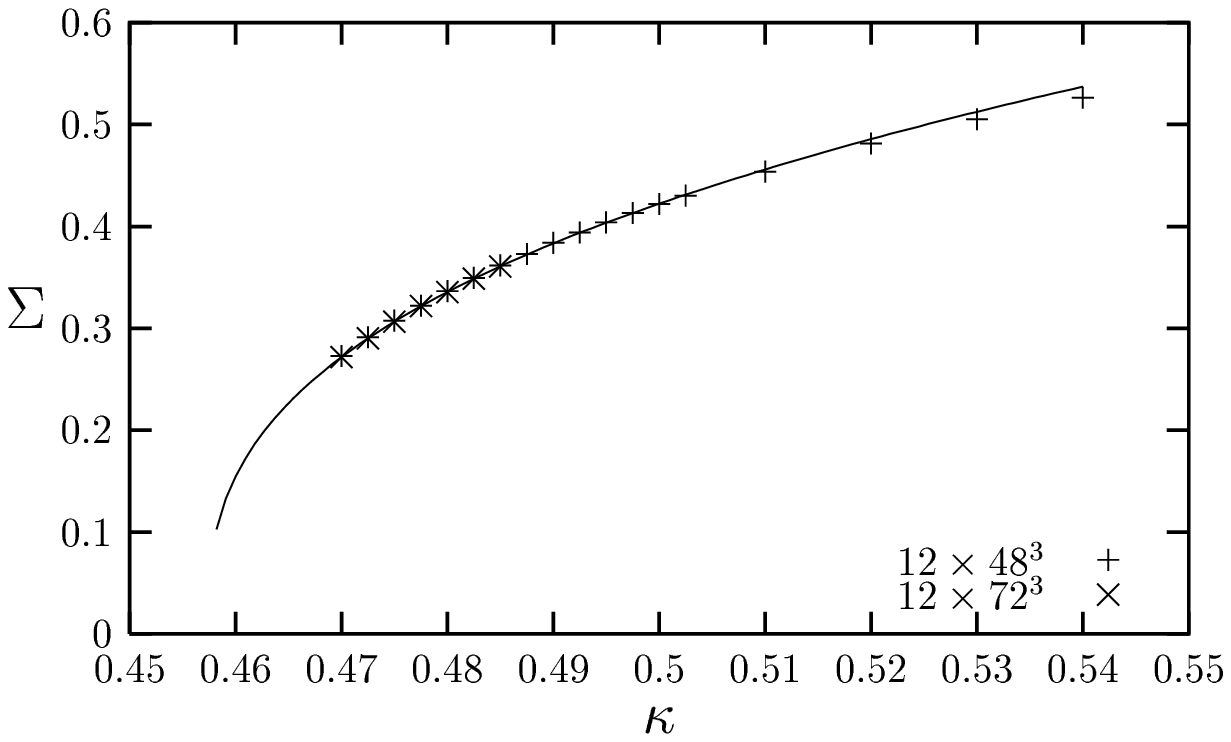}
\caption{Order parameter vs. coupling for the $(3+1)d$ $O(3)$ model
measured on $12 \times 48^3$ and $12 \times 72^3$ lattices.
The curve represents the power law fit to the data for $\kappa \leq 0.4875$.}
\label{order.papram_O3}
}

We also performed simulations of the $O(3)$ model on lattices with 
sizes $12 \times 48^3$ and $12 \times 72^3$ and we observed a similar scenario 
to the one described above for the $O(2)$ model. Again, 
we generated approximately $30,000$ to $50,000$ configurations for each 
value of $\kappa$ on both lattice sizes.
The values of $\beta_{\rm mag}$ extracted from fits near the transition are 
very close to recent numerical results which predicted $\beta_{\rm mag}^{\rm 3d}=0.3689(3)$
\cite{hasenbusch02}. 
The discrepancy in the values of $\beta_{\rm mag}$ extracted from the simulations on the two different 
lattice sizes and the difference in the value of $\chi^2/{\rm d.o.f.}$ could be attributed 
to finite size effects on the smaller lattice. 
As expected the finite size effects in the $O(3)$ model are larger than in the 
$O(2)$ model, because it has twice as many Goldstone bosons.
Again by monitoring the quality of the fit and the deviation of $\beta_{\rm mag}$ 
from $\beta_{\rm mag}^{\rm 3d}$ we found that the crossover into the $3d$ scaling region 
occurs at $\kappa_{\rm cross} \approx 0.4975$.
The values of the spatial and temporal correlation lengths measured 
on a $12 \times 24^3$ lattice at the crossover coupling 
$\kappa_{\rm cross} = 0.4975$ are
$\xi^{s}_{\rm cross}=0.92(3)$ and $\xi^{t}_{\rm cross}=0.90(2)$.
Given the uncertainty in the measurement of the $\kappa_{\rm cross}$, 
we conclude that the crossover in the $O(2)$ and $O(3)$ models occurs
at approximately the same value of the correlation length. 

\TABLE
{
\centering
\caption{Lattice sizes and statistics at different values of $\kappa$ for the 
$(2+1)d$ Ising model simulations.}
\label{t4}       
\setlength{\tabcolsep}{2.5pc}
\begin{tabular}{|lll|}
\hline
range of $\kappa$ & $L_s$ & statistics \\
\hline
0.22425 - 0.22800 & 240   & 100,000  \\
0.22404 - 0.22417 & 480   & 300,000  \\
0.22400, 0.22404  & 720   & 500,000  \\
\hline
\end{tabular}
}

\FIGURE
{
\centering
\includegraphics[width=11cm]{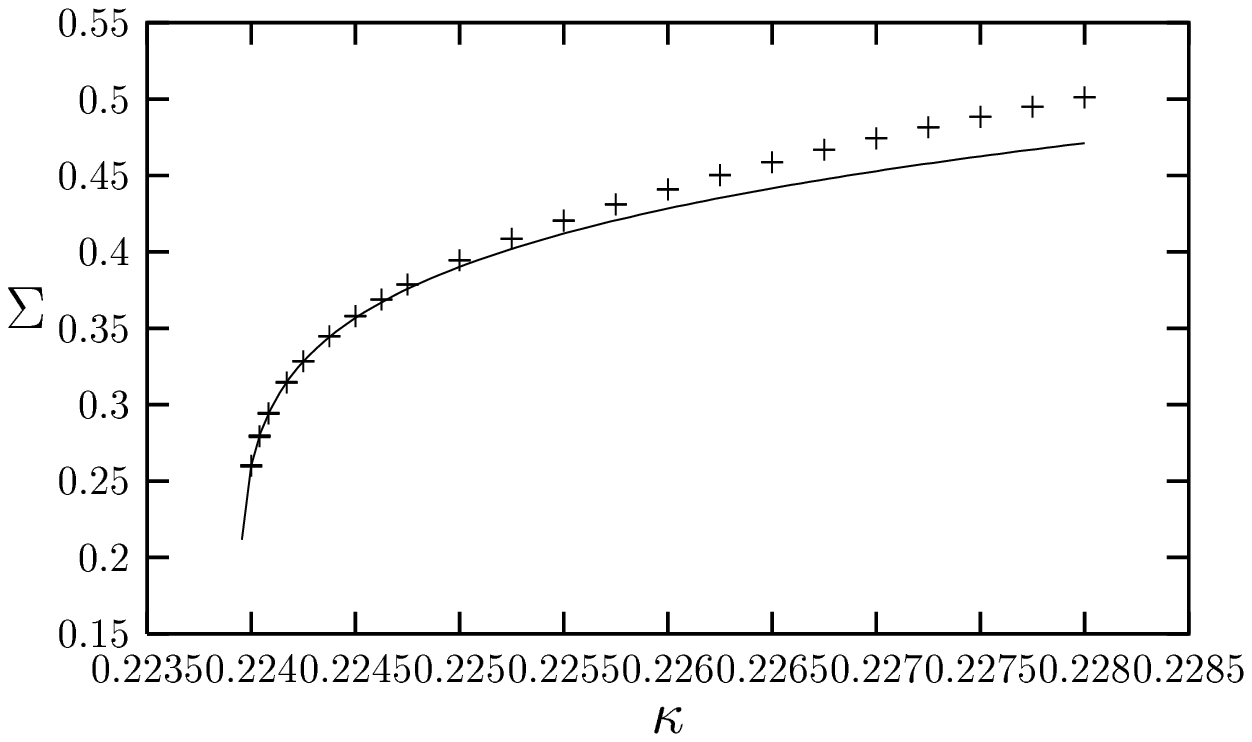}
\caption{Order parameter vs. coupling for $(2+1)d$ Ising model.
The curve represents the power law fit to the data for $\kappa \leq 0.22400$.}
\label{3d.ising}
}

We also performed simulations of the $(2+1)d$ Ising model. The results 
for the order parameter are shown in 
Fig.~\ref{3d.ising}. In this case, because the dimensionality is lower, 
the infrared 
fluctuations are stronger than in the $(3+1)d$ model with the same symmetry. 
Therefore, in order to suppress finite size effects 
we simulated the model on lattices with larger asymmetry ratios than 
in the $(3+1)d$ case. The values of $L_s$ and the amount of statistics 
generated at various values of $\kappa$ are shown in Table \ref{t4}.
By fitting the data with the 
four smallest values of $\kappa$ to the standard scaling relation we extracted 
$\kappa_c=0.22394(1)$ and
$\beta_{\rm mag}=0.138(8)$, which is close to the $2d$ Ising exponent $\beta_{\rm mag}^{\rm 2d}=0.125$.
We conclude that the crossover into the $2d$ scaling region is at $\kappa_{\rm cross} \approx 0.22417$.
The value of the spatial correlation length at $\kappa_{\rm cross}$ 
measured on a $12\times240^2$ lattice by generating $5.5$ million configurations 
is  $\xi^{s}_{\rm cross} = 25.0(2.0)$. 
Unfortunately, the large statistical fluctuations in the data set did not allow us to measure
the temporal correlation length.

Next, we present results in the symmetric phase of the $O(2)$ model. 
We measured the improved estimators for the susceptibility (see eq.~(\ref{suscimp})),
for the exponential correlation 
length $\xi_{\rm exp}$ extracted from the asymptotic decay of $C(t)_{\rm imp}$ 
(see eq.~(\ref{xiexpimp}))
and for the second moment correlation length $\xi_{\rm 2nd}$ (see eq.~(\ref{xi2imp})). 
It is well-known that in the high temperature phase of this model $L_s/\xi \approx 7$ is
sufficient to give thermodynamic limit results.
In order to detect finite size effects, 
we performed simulations at $\kappa=0.2930$ on lattices with $L_s=24$ and $36$  and 
we found that in both cases the values of the susceptibility  
and the correlation lengths agree within errors,
implying that finite-size effects are under control. 
In most cases we kept $L_s/\xi > 9$ by increasing $L_s$ from $24$ to $96$ as
we approached the critical coupling with constant $L_t=12$. 
For each value of the coupling we generated 
approximately $10^7$ configurations. We also checked that the deviations 
between $\xi_{\rm exp}$ and $\xi_{\rm 2nd}$
are less than $0.5\%$ for all the values of $\kappa$.
An interesting observation is that $\xi^{t}$ and 
$\xi^{s}$ remain equal for 
all values of $\kappa$ studied in this work. 
The value of $\kappa$ which is closest to $\kappa_c$ is $0.3015$ and $\xi=14.18(7)$
at this value of the coupling.
This observation implies that in the symmetric phase the quantum fluctuations remain more important than the 
thermal fluctuations up to relatively large values of $\xi/L_t$.

\TABLE
{
\centering
\caption{Values of $\gamma$ and $\nu$ measured in the symmetric phase of the $(3+1)d$ O(2) model in 
different ranges of $\kappa$ and $\xi^s$.}
\label{t5}       
\setlength{\tabcolsep}{2.0pc} 
\begin{tabular}{|llll|}
\hline 
range of $\kappa$ & range of $\xi^s$ & $\gamma$ & $\nu$ \\
\hline
0.2800 - 0.2970 & 1.691(2) - 4.003(7)  & 1.11(1)  & 0.602(4) \\
0.2990 - 0.3015 & 5.33(3) - 14.18(7)   & 1.18(1)  & 0.59(1)  \\
\hline
\end{tabular}
}

\FIGURE
{
\centering
\includegraphics[width=11cm]{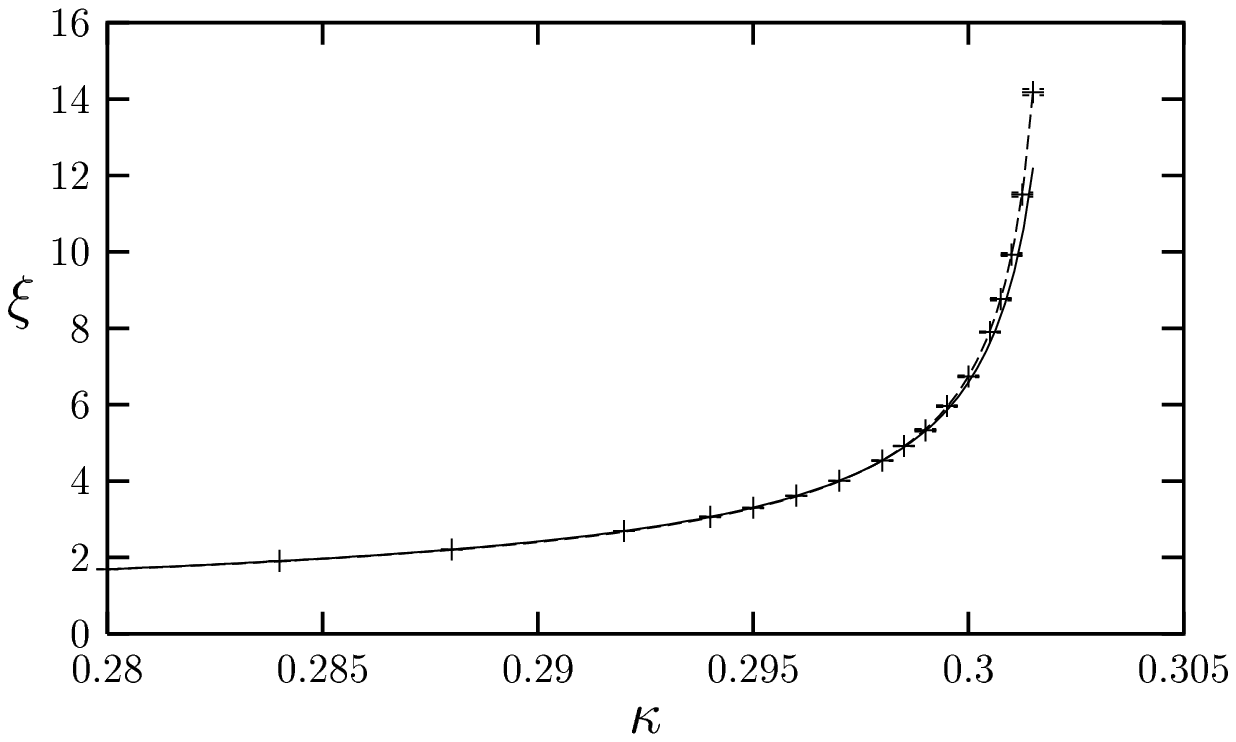}
\caption{Correlation length vs. coupling measured in the symmetric
phase of the $(3+1)d$ $O(2)$ model. The continuous line represents the
fit to the data for $0.2800 \leq \kappa \leq 0.2970$ and the dashed line
represents the fit for $0.2990 \leq \kappa \leq 0.3015$.}
\label{correl_O2_symmetric}
}

\FIGURE
{
\centering
\includegraphics[width=11cm]{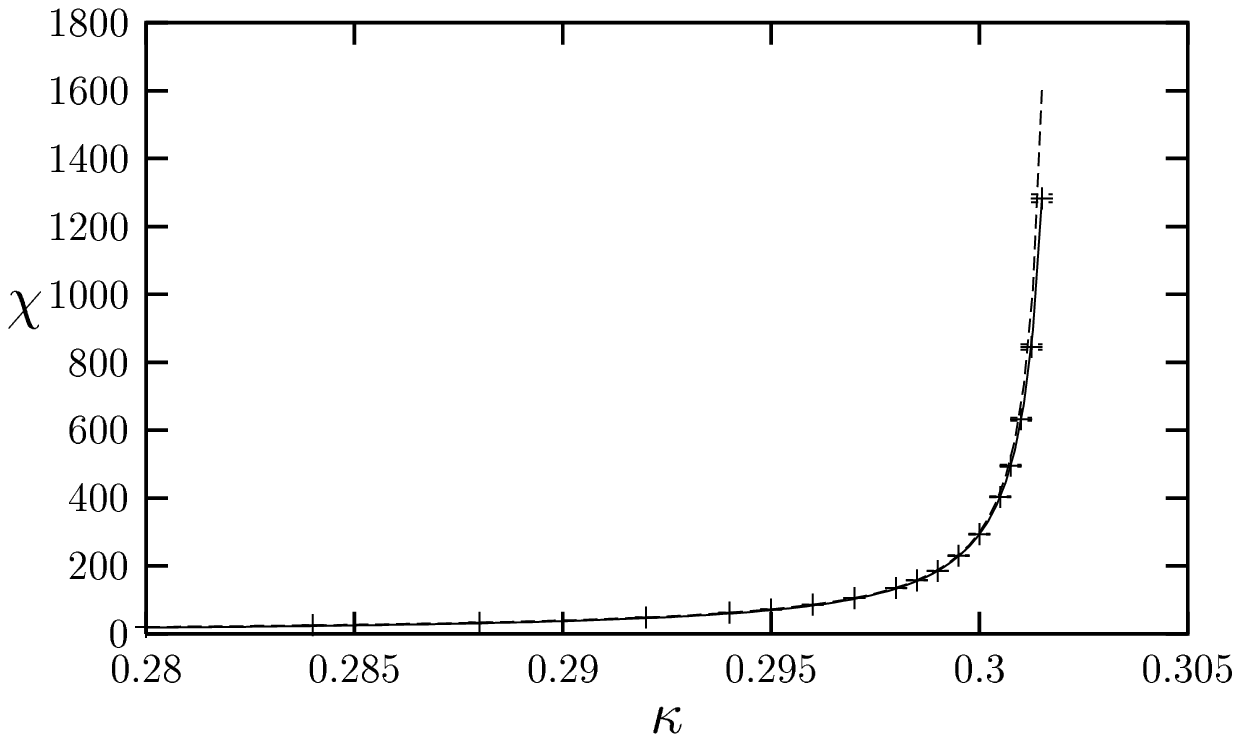}
\caption{Susceptibility vs. coupling measured in the symmetric
phase of the $(3+1)d$ $O(2)$ model. The continuous line represents the
fit to the data for $0.2800 \leq \kappa \leq 0.2970$ and the dashed line
represents the fit for $0.2990 \leq \kappa \leq 0.3015$.}
\label{suscept_O2_symmetric}
}

We also measured the exponents $\gamma$ and $\nu$ in the $O(2)$-symmetric 
phase by fitting our data 
to the standard scaling laws
\begin{equation}
\chi=B(\kappa_c-\kappa)^{-\nu}; \;\;\;\; \xi=C(\kappa_c-\kappa)^{-\gamma}.
\end{equation}
The results for the fits in different scaling windows are shown in Table \ref{t5}.
The data and the fitting curves are shown in 
Figs.~\ref{correl_O2_symmetric} and \ref{suscept_O2_symmetric}. 
The exponents measured are significantly different from    
recent predictions of the three-dimensional critical exponents $\gamma^{\rm 3d}=1.3177(5)$
and $\nu^{\rm 3d}=0.67155(27)$ \cite{hasenbusch01}.
Our results are consistent however, with the observation mentioned above that in the 
symmetric phase the thermal fluctuations
do not influence decisively the critical properties of the model even up to values of the correlation 
length that are much larger than the value of $\xi_{\rm cross}$ in the broken 
phase.

\section{Conclusions} 
We presented results from numerical simulations of the $O(N)$ sigma model 
at finite temperature for $N=1,2,3$ in four dimensions and $N=1$ in three
dimensions. We investigated the phenomenon of dimensional reduction
and studied the dependence of the width of the critical region on the number of field components $N$ and
on whether the $O(N)$ symmetry is spontaneously broken or restored. 
Contrary to the 
naive geometric scenario which requires that the condition $\xi \gg L_t$ must be satisfied 
in order to reach the $d-1$ scaling region, 
we showed that in the low temperature phase the correlation length of the sigma meson doesn't have to be
much larger than $L_t$ in order to reach the dimensionally reduced critical region.
Our results imply that the nonzero 
Matsubara modes do not affect the critical singularities but they can affect non-universal
aspects of the transition such as the width of the scaling region.
We measured the correlation length of the sigma meson in the temporal and spatial directions 
at the crossover into the
$d-1$ scaling region and
we found that for $\frac{\Lambda}{T}=L_t=12$ in the case of the $O(2)$ model 
$\xi^{t}_{\rm cross}=0.94(2)$ and $\xi^{s}_{\rm cross}=0.97(3)$ and for the $O(3)$ model 
$\xi^{t}_{\rm cross}=0.90(2)$ and $\xi^{s}_{\rm cross}=0.92(3)$. We should note however, 
that the deviations from the $3d$ scaling laws at these relatively small correlation 
lengths should also be affected by discretization effects.  
In the case of the Ising model we measured $\xi^{t}_{\rm cross}=6.5(4)$ 
and $\xi^{s}_{\rm cross}=8.8(5)$,
implying that the critical region is much wider in the presence of the  
infrared fluctuations due to the Goldstone bosons. 
The critical regions
for $N=2$ and $3$ have approximately the same width and therefore we don't expect a 
significant change in the width of the scaling region for $N >3$. 
The fact that for the continuous symmetries $\xi^{s}_{\rm cross} \approx \xi^{t}_{\rm cross}$
implies that the thermal fluctuations are still very small at the crossover temperature.

It is not yet known whether the dimensional reduction scenario is the correct description of
the QCD thermal phase transition. If it is, then the thermal transition with massless
quarks may belong to the $O(4)$ universality class, whereas when the quarks are massive
the tricritical point in the $(\mu,T)$ plane becomes critical and it must belong to the 
$3d$ Ising universality class. According to our results the latter must have a much thinner
scaling region than the massless quarks transition.

We also showed that the critical region of the $(3+1)d$ $O(2)$ model  in the symmetric 
phase is much thinner than in the broken phase. 
Despite the fact that we simulated the model very close to the critical point and went 
up to correlation lengths $\approx 14$, the values of the measured exponents are not consistent
with the $3d$ exponents; they are rather effective exponents with values between mean field exponents
and $3d$ exponents.

We also studied the symmetry restoration transition in the $(2+1)d$ Ising model.
This model has a well-defined interacting continuum limit and its behavior
near $T_c$ is particularly hard to explore analytically, because infrared
divergences get worsened in three dimensions. We showed that in order to reach 
the $2d$ scaling region on lattices with temporal extent $L_t=12$ the spatial 
correlation length $\xi^{s}=25.0(2.0)$ and in order to minimize finite size effects
the simulations near $T_c$ were performed on lattices with large spatial extents 
that are larger by almost an order of magnitude than in the four-dimensional case.
In this lower dimensionality we studied only $N=1$, because
according to the Coleman-Mermin-Wagner theorem \cite{coleman} for any continuous symmetry in two
spatial dimensions 
the infrared divergent fluctuations wash out the order parameter 
and the symmetry is manifest for $T>0$.

In the future we would like to study the width of the scaling region of the $O(N)$ linear
sigma model and compare with the results from the nonlinear sigma model presented in this
paper. It was shown in the past that the dimensional reduction scenario
is the correct description of the chiral phase transition in four-fermion models \cite{strouthos}, 
because the infrared region  
is dominated by the zero Matsubara mode of the bosonic field and the nonzero
bosonic and fermionic modes decouple. 
Simulations of the $(3+1)d$ four-fermion model with $Z_2$ and $SU(2) \times SU(2)$ chiral symmetries are underway in order to study the dependence of the
width of the critical region on the degree of chiral symmetry of the model.

\acknowledgments
Discussions with Simon Hands and John Kogut are greatly appreciated. 
Costas Strouthos was supported by a Leverhulme Trust grant. Ioannis Tziligakis
was partially supported by the Research Board at the University of Illinois, 
RES BRD KOGUT J 1-1-28212. Funds provided by this grant were used to 
purchase a dual processor Pentium IV,2.2 Ghz computer, on which the simulations
were carried out. Ioannis Tziligakis would also like to thank Costas Papanikolas and the Institute
for Accelerator Systems and Applications (IASA) at the University of Athens, Greece
for their hospitality in the fall of 2002.

\renewcommand{\theequation}{A-\arabic{equation}}
\setcounter{equation}{0}  

\end{document}